\documentclass[reprint, aps, nofootinbib, prb]{revtex4-1}
\usepackage{amsmath} \usepackage{graphicx} \usepackage{dcolumn}
\usepackage{bm} 
\usepackage{amssymb}
\usepackage{pstricks}
\usepackage{subfigure}

\begin{document}

\newlength{\figurewidth}
\setlength{\figurewidth}{\columnwidth}

\newcommand{\prtl}{\partial}
\newcommand{\la}{\left\langle}
\newcommand{\ra}{\right\rangle}
\newcommand{\dla}{\la \! \! \! \la}
\newcommand{\dra}{\ra \! \! \! \ra}
\newcommand{\we}{\widetilde}
\newcommand{\smfp}{{\mbox{\scriptsize mfp}}}
\newcommand{\smp}{{\mbox{\scriptsize mp}}}
\newcommand{\sph}{{\mbox{\scriptsize ph}}}
\newcommand{\sinhom}{{\mbox{\scriptsize inhom}}}
\newcommand{\sneigh}{{\mbox{\scriptsize neigh}}}
\newcommand{\srlxn}{{\mbox{\scriptsize rlxn}}}
\newcommand{\svibr}{{\mbox{\scriptsize vibr}}}
\newcommand{\smicro}{{\mbox{\scriptsize micro}}}
\newcommand{\scoll}{{\mbox{\scriptsize coll}}}
\newcommand{\sattr}{{\mbox{\scriptsize attr}}}
\newcommand{\sth}{{\mbox{\scriptsize th}}}
\newcommand{\sauto}{{\mbox{\scriptsize auto}}}
\newcommand{\seq}{{\mbox{\scriptsize eq}}}
\newcommand{\teq}{{\mbox{\tiny eq}}}
\newcommand{\sinn}{{\mbox{\scriptsize in}}}
\newcommand{\suni}{{\mbox{\scriptsize uni}}}
\newcommand{\tin}{{\mbox{\tiny (in)}}}
\newcommand{\tout}{{\mbox{\tiny (out)}}}
\newcommand{\scr}{{\mbox{\scriptsize cr}}}
\newcommand{\tstring}{{\mbox{\tiny string}}}
\newcommand{\sperc}{{\mbox{\scriptsize perc}}}
\newcommand{\tperc}{{\mbox{\tiny perc}}}
\newcommand{\sstring}{{\mbox{\scriptsize string}}}
\newcommand{\stheor}{{\mbox{\scriptsize theor}}}
\newcommand{\sGS}{{\mbox{\scriptsize GS}}}
\newcommand{\sBP}{{\mbox{\scriptsize BP}}}
\newcommand{\sNMT}{{\mbox{\scriptsize NMT}}}
\newcommand{\sbulk}{{\mbox{\scriptsize bulk}}}
\newcommand{\tbulk}{{\mbox{\tiny bulk}}}
\newcommand{\sXtal}{{\mbox{\scriptsize Xtal}}}
\newcommand{\sliq}{{\text{\tiny liq}}}

\newcommand{\smin}{\text{min}}
\newcommand{\smax}{\text{max}}

\newcommand{\saX}{\text{\tiny aX}}
\newcommand{\slaX}{\text{l,{\tiny aX}}}

\newcommand{\svap}{{\mbox{\scriptsize vap}}}
\newcommand{\sjam}{J}
\newcommand{\Tm}{T_m}
\newcommand{\sTS}{{\mbox{\scriptsize TS}}}
\newcommand{\sDW}{{\mbox{\tiny DW}}}
\newcommand{\cN}{{\cal N}}
\newcommand{\cB}{{\cal B}}
\newcommand{\br}{\bm r}
\newcommand{\be}{\bm e}
\newcommand{\cH}{{\cal H}}
\newcommand{\cHlt}{\cH_{\mbox{\scriptsize lat}}}
\newcommand{\sthermo}{{\mbox{\scriptsize thermo}}}

\newcommand{\bu}{\bm u}
\newcommand{\bk}{\bm k}
\newcommand{\bX}{\bm X}
\newcommand{\bY}{\bm Y}
\newcommand{\bA}{\bm A}
\newcommand{\bb}{\bm b}

\newcommand{\lintf}{l_\text{intf}}

\newcommand{\DV}{\delta V_{12}}
\newcommand{\sout}{{\mbox{\scriptsize out}}}
\newcommand{\dv}{\Delta v_{1 \infty}}
\newcommand{\dvin}{\Delta v_{2 \infty}}

\newcommand*\xbar[1]{%
  \hbox{%
    \vbox{%
      \hrule height 0.5pt 
      \kern0.5ex
      \hbox{%
        \kern-0.1em
        \ensuremath{#1}%
        \kern-0.1em
      }%
    }%
  }%
}

\newcommand{\cV}{{\cal V}}

\def\Xint#1{\mathchoice
   {\XXint\displaystyle\textstyle{#1}}%
   {\XXint\textstyle\scriptstyle{#1}}%
   {\XXint\scriptstyle\scriptscriptstyle{#1}}%
   {\XXint\scriptscriptstyle\scriptscriptstyle{#1}}%
   \!\int}
\def\XXint#1#2#3{{\setbox0=\hbox{$#1{#2#3}{\int}$}
     \vcenter{\hbox{$#2#3$}}\kern-.5\wd0}}
\def\ddashint{\Xint=}
\def\dashint{\Xint-}
\title{A mechanism for reversible mesoscopic aggregation in liquid
  solutions}

\author{Ho Yin Chan} \affiliation{Department of Chemistry, University
  of Houston, Houston, TX 77204-5003} \affiliation{Department of
  Physics, University of Houston, Houston, TX 77204-5005}

\author{Vassiliy Lubchenko} \email{vas@uh.edu} \affiliation{Department
  of Chemistry, University of Houston, Houston, TX 77204-5003}
\affiliation{Department of Physics, University of Houston, Houston, TX
  77204-5005}

\date{\today}

\begin{abstract}

  We show systematically that a steady-state ensemble of mesoscopic
  inclusions of a solute-rich fluid can emerge in liquid solutions
  well outside the region of stability of the solute-rich
  phase. Unanticipated by conventional treatments, this type of
  reversible aggregation nonetheless can take place if the solute
  molecules bind transiently with each other to form long-lived
  complexes. The binding causes kinetic stabilization of inclusions of
  the solute-rich phase---within a substantial size range---so as to
  render the critical size for nucleation of the inclusions
  finite. Individual droplets nucleate and grow until they become
  mechanically unstable because of a concomitant drop in the internal
  pressure, the latter drop caused by the thermodynamic metastability
  of the solute-rich phase. At the same time, the {\em ensemble} of
  the droplets is steady-state on long times. In a freshly prepared
  solution, the ensemble is predicted to evolve similarly to the
  conventional Ostwald ripening, during which larger droplets grow at
  the expense of smaller droplets. The present mechanism is proposed
  to underlie the puzzling mesoscopic clusters observed in solutions
  of proteins and other molecules.

\end{abstract}

\maketitle


Spatially and chemically heterogeneous systems are of prime
significance in the context of both man-made processes, such as
self-assembly and nano-particle manufacturing, and naturally occurring
systems, such as membrane-less organelles.~\cite{Elbaum-Garfinkle7189,
  BanjadeE6426, Shineaaf4382, 0034-4885-81-4-046601} In many cases,
spatially inhomogeneous solutions are not steady-state but, instead,
only long-lived; examples include micelle assemblies and various
lamellar phases.  This is in contrast with equilibrated liquid
solutions, which must be spatially uniform.  In addition, a phase of
matter will eventually reach macroscopic dimensions, if
thermodynamically stable.  If metastable, a phase will not be
typically observed in equilibrium: Nucleation of a metastable phase is
an uphill process, free energy-wise, owing to the bulk free energy
cost and the mismatch penalty between the majority an minority
phase.~\cite{GibbsV1, LLstat} Thus in equilibrium, heterogeneities
must be either of macroscopic dimensions or not present at all.

It then comes as a surprise that equilibrated solutions of several
proteins must host mesoscopically-sized inclusions of what seems to be
a distinct, protein-rich phase of fluid
consistency;~\cite{GlikoJACS2005, Georgalis1999,
  doi:10.1021/jp068827o, PVL, LLVFPostwaldRipening, YEARLEY20141763,
  SleutelE546, doi:10.1021/acs.cgd.6b01826, Gliko2005} these
inclusions are often called {\em mesoscopic
  clusters}. Cluster-containing solutions are stable on time scales of
a few months.~\cite{LLVFPostwaldRipening} In systems studied so far,
the mesoscopic clusters contain a small fraction of the solute, less
than $10^{-3}$, and thus do not affect the appearance of the solution;
common methods of detection include dynamic light scattering, direct
tracking using fluorescence, and also atomic force microscopy.  In
addition to solutions of many proteins, mesoscopic clusters have been
recently observed in solutions of relatively simple molecules, viz.,
the pharmaceutical olanzapine.~\cite{doi:10.1021/acs.cgd.7b01299}

The mesoscopic clusters are important for many reasons: They serve as
essential nucleation sites for solid protein aggregates such as sickle
cell anemia fibers~\cite{UZUNOVA20101976, C2FD20058A} and protein
crystals.~\cite{GlikoJACS2005, SleutelE546, Yamazaki2154,
  doi:10.1063/1.1943389} Thus by deliberately inducing the formation
of clusters, one can seed formation of solid aggregates of interest in
applications. Equally important seems the fact that the clusters form
an ensemble of objects whose size is narrowly distributed around a
steady-state value. This may provide a separate avenue for making
mesoscopically sized particles or gels in industrially relevant
quantities. On the more fundamental side, we believe that the
existence of mesoscopic clusters suggest a tantalizing possibility
that the precursors to living cells were not encased in membranes but,
instead, were more like the so called membrane-less
organelles. Differing from the surrounding cytoplasm chemically,
membrane-less organelles~\cite{Elbaum-Garfinkle7189, BanjadeE6426,
  Shineaaf4382, 0034-4885-81-4-046601} essentially serve as cell's
chemical reactors; the lack of a membrane provides for ready exchange
of reactants and products with the cytoplasm.  In view of the
continuously growing number of cluster sightings, it stands to reason
than the clusters are more common than one might think, but are not
detected more frequently either because they are a kinetic
intermediate to a more stable phase or simply for the lack of trying.

The mesoscopic clusters are not micelle-like objects. This is
evidenced by the fact that the mole fraction of the clusters increases
with the concentration of the solute; the value of the mole fraction
is consistent with estimates of the free energy cost of creating bulk
solute-rich liquid.~\cite{PVL} At the same time, the typical {\em
  size} of an individual cluster does not sensitively depend on the
solute concentration. This is in contradistinction with macroscopic
phases, which respond to changing conditions by evolving in size until
the solution is again saturated. Still in one particular way, the
clusters in freshly prepared solutions behave similarly to macroscopic
phases: Well before its steady-state value is reached, the typical
cluster size depends on time~\cite{LLVFPostwaldRipening} in a way
reminiscent of Ostwald ripening.~\cite{LifshitzSlyozov1961, WagnerOR,
  Bray, PhysRevA.20.595}
   
The lack of dependence of the steady-state cluster size on the solute
concentration in the bulk solution suggests an additional,
molecular-level process is at work.  Pan et al.~\cite{PVL} proposed
that this additional process involves the formation and decay of a
solute-containing species, call it the ``complex.''  In this
mechanism, the solute-rich phase is assumed to be rich in the
complex. The complex would have to have a relatively long
lifetime---of the order milliseconds for protein solutions---and could
be a dimer or a higher-order oligomer made of the monomers.  The
cluster size $R$ is essentially determined by the distance the complex
can diffuse before it decays:
\begin{equation} \label{RDk} R \approx \sqrt{D_c/k}.
\end{equation}
where $k$ is the decay rate of the complex and $D_c$ its
diffusivity. The lengthscale $R$ emerges self-consistently as a result
of solving a set of reaction-diffusion equations applicable near the
cluster edge.~\cite{PVL} Inside the cluster, the equations become
however invalid and, furthermore, produce unphysical singularities.


Lutsko and Nicolis~\cite{C5SM02234G} (LN) extended the Pan et al.'s
treatment to explicitly include particle-particle interactions using a
standard approximation of the theory of liquids. These authors
concluded that the resulting reaction-diffusion equations allow for a
stationary solution in the form of {\em stable individual clusters}, a
startling result indeed. Note such a stationary solution would not
allow for Ostwald-like ripening but, instead, would exhibit simpler,
exponential kinetics for the relaxation of cluster size. At the same
time, the only known mechanism for bona fide Ostwald ripening---which
may or may not apply to the clusters---requires that droplets
surrounded by under-saturated solution evaporate while droplets
surrounded by over-saturated solution grow
indefinitely.~\cite{LifshitzSlyozov1961, WagnerOR} Perhaps fittingly,
Lutsko~\cite{0953-8984-28-24-244020} concluded in a subsequent
analysis that realistically accounting for the variability of the
kinetics depending on the solute concentration would disrupt the
complexation mechanism put forth in Ref.~\onlinecite{PVL}, after all.

Here we present a fully internally-consistent calculation
demonstrating that the complexation scenario can, in fact, lead to the
emergence of a {\em metastable} minority phase that is fragmented into
inclusions of substantial yet non-macroscopic size, or ``clusters.''
In contrast with the conclusions of the LN study, {\em individual}
clusters are never stable in the present mechanism. Once nucleated,
the clusters grow precipitously until they become mechanically
unstable. Because the minority phase is metastable, its bulk pressure
is automatically lower than that in the bulk phase. For sufficiently
small clusters, this deficit of pressure is offset by the excess,
Laplace pressure due to the curvature of the interface between the
minority and majority phase. Yet this curvature-caused pressure
decreases as $1/R$ with the cluster size $R$. Thus for sufficiently
large droplets, the hydrostatic pressure on the inside becomes lower
than on the outside; this eventually causes a mechanical instability
toward caving or necking. Thus in an equilibrated solution, the
droplets nucleate, grow, and decay at a steady rate leading to a
steady-state ensemble of clusters but not steady-state individual
clusters.

The formation of the transient complexes serves to effectively provide
partial, kinetic stabilization of the minority phase but on
lengthscales comparable or less than the distance a complex can travel
before it decays. (As in Ref.~\onlinecite{PVL}, the complexation
mechanism requires that the solute-rich phase is rich in the complex.)
Thus the question of whether the clusters could nucleate is the
question of whether microscopic parameters could conspire to make the
critical size for cluster nucleation shorter than the kinetic length
from Eq.~(\ref{RDk}). Here we show that, indeed, there is a
substantial range of microscopic parameters for which the answer to
this question is affirmative. At the same time, the cluster size at
the mechanical stability edge does not change much within that
parameter range, which is consistent with the observed behavior in
protein solutions.  Non-withstanding the kinetic character of the
effective bulk stabilization of the minority phase, due to the complex
formation, cluster nucleation shares an important aspect with
nucleation of a {\em stable} minority phase: The effective value of
saturated vapor pressure still depends on the cluster size. Thus one
expects the clusters should exhibit Ostwald-like ripening at
sufficiently early times, again consistent with observation.  Finally,
may types of solutes exhibit a propensity for the formation of
transient complexes, even if short-lived. Thus we predict mesoscopic
clusters should be observed commonly even if not universally.

\section*{Setup of the calculation and results}

For concreteness, we assume that the solute-containing complex is a
dimer. The coordinate-dependent concentrations of the solute (``the
monomer'') and the complex (``the dimer'') are denoted with $n_1$ and
$n_2$, respectively. The corresponding reaction-diffusion scheme is
\begin{align} \label{particleCons_nonlin} \begin{array}{ll}
    \dot{n}_1 &= -\nabla \bm j_1 - k_1 n_1^2 + 2 k_2 n_2 \\  \vspace{-2mm} \\
    \dot{n}_2 &= -\nabla \bm j_2 + \frac{1}{2} k_1 n_1^2 - k_2 n_2,
  \end{array}
\end{align}
where $\bm j_i$ is the flux of species $i$, $k_1$ the (bi-molecular)
rate of binding of the monomer to itself, and $k_2$ the dissociation
rate of the dimer. Strictly speaking, the reaction terms in
Eq.~(\ref{particleCons_nonlin}) should be written using the
activities, not concentrations; we will return to this notion shortly.

The transport for each species is overdamped at the conditions of
interest and thus obeys the usual Fick's law:~\cite{Bray}
\begin{equation} \label{Fick} \bm j_i = -\widetilde D_i \nabla \mu_i,
\end{equation}
where $\widetilde D_i$ is the self-diffusivity of species $i$ and
$\mu_i$ is its chemical potential. To include off-equilibrium
situations in the treatment, we allow both the chemical potentials and
concentrations of the monomer and dimer to be coordinate-dependent.
The local value of the chemical potential, by construction, is the
free energy cost of adding a particle to the system at the locale in
question:
\begin{equation} \label{eq:mu_definition} \mu_i(\br) = \frac{\delta
    F}{\delta n_i(\br)},
\end{equation}
where $F$ is the total free energy of the system and $\delta/\delta
n_i$ is the functional derivative with respect to $n_i$.~\cite{Bray,
  Goldenfeld, L_AP}

It is guaranteed~\cite{PhysRev.137.A1441} that there is a unique free
energy density functional that is optimized by the equilibrium density
profiles.  Irrespective of the detailed form of approximation used for
the density functional,
Eqs.~(\ref{particleCons_nonlin})-(\ref{eq:mu_definition}) provide a
complete description of transport and inter-conversion of the monomer
and dimer. This description automatically obeys detailed balance and
conservation laws.

As a practical matter, one uses an approximate form for the free
energy functional such as the venerable
Landau-Ginzburg-Cahn-Hilliard~\cite{CahnHilliard} functional, which we
employ here as well:
\begin{align} \label{eq:free_energy} F &= \int \left[
    \frac{\kappa_1}{2} \left(\nabla n_1 \right)^2 + \frac{\kappa_2}{2}
    \left(\nabla n_2 \right)^2 + \cV \left(n_1, n_2\right) \right] d^3
  \br.
\end{align}
The latter functional affords one a quantitative description not too
close to criticality.~\cite{Goldenfeld} We assume that the
monomer-dimer-buffer mixture can have two distinct liquid phases, one
monomer-rich and the other dimer-rich.  The bulk portion of the
corresponding free energy functional thus has two distinct minima,
which makes the solution of
Eqs.~(\ref{particleCons_nonlin})-(\ref{eq:mu_definition}) difficult in
the interfacial region. These difficulties can be efficiently
addressed,~\cite{CL_LG} as we detail in the Methods, by adopting
parabolic free energy profiles everywhere within individual
phases. Thus the bulk free energy of the mixture is set, by
construction, at
\begin{align} \label{ftilde2} \cV\left(n_1, n_2\right) &=\min_\alpha
  \left[ g^{(\alpha)} \right. \\ &+ \left.
    \frac{m_1^{(\alpha)}}{2}\left(n_1-n_{1,b}^{(\alpha)}\right)^2 +
    \frac{m_2^{(\alpha)}}{2}\left(n_2-n_{2,b}^{(\alpha)}\right)^2
  \right] \nonumber,
\end{align}
where $\alpha$ labels the phase: $\alpha = \text{m}$ for the
monomer-rich, and $\alpha = \text{d}$ for the dimer-rich solution.
The quantity $n_{i, b}^{(\alpha)}$ denotes the equilibrium bulk value
of the concentration of species $i$ in phase $\alpha$. These are
connected with the rate constants according to $k_1^{(\alpha)}
(n_{1,b}^{(\alpha)})^2 = 2 k_2^{(\alpha)} n_{2,b}^{(\alpha)}$.  The
coefficients $m_i^{(\alpha)}$ reflect the free energy penalty for
density fluctuations and are proportional to the inverse osmotic
compressibility.

In the present treatment of thermodynamics ($\kappa_i$, $m_i$),
transport ($\widetilde{D}_i$), and chemical transformation ($k_i$), we
are performing a quadratic expansion around the bulk equilibrium state
for each individual phase. This allows us to use
concentration-independent coefficients $\kappa_i$, the diffusivities
$\widetilde{D}_i$, and rate coefficients $k_i$, while writing down the
kinetic terms in Eqs.~(\ref{particleCons_nonlin}) in terms of
concentrations, not activities.  Assumed to constant within individual
phases, these coefficients generally differ between distinct
phases. Clearly, the variation of the parameters {\em between} the
phases, not within individual phases, is the most important
effect. The present approach captures this effect. We note that the
four diffusivities---there are two species and two phases---are not
independent. For internal consistency, one must set
$\widetilde{D}_1^\text{(m)}/\widetilde{D}_2^\text{(m)} =
\widetilde{D}_1^\text{(d)}/\widetilde{D}_2^\text{(d)}$, see Methods
for details.

Additional computational difficulties are caused by the presence of
the non-linear term $k_2 n_1^2$ in Eq.~(\ref{particleCons_nonlin}).
We have numerically solved the resulting non-linear differential
equations for several realizations of parameters---to be discussed in
due time---however the majority of the calculations were performed for
a linearized version of Eq.~(\ref{particleCons_nonlin}) so that the
interconversion between the two species is effectively a first order
reaction:
\begin{align} \label{particleCons} \begin{array}{ll} \dot{n}_1 &=
    -\nabla \bm j_1 - k_1 n_1
    + k_2 n_2 \\  \vspace{-2mm} \\
    \dot{n}_2 &= -\nabla \bm j_2 + k_1 n_1 - k_2 n_2,
\end{array}
\end{align}
where $k_1 n_{1, b} = k_2 n_{2, b}$ in each phase.  Note that in going
from Eq.~(\ref{particleCons_nonlin}) to Eq.~(\ref{particleCons}) we
have made a variable change, for convenience. As a result of this
change, the chemical potential of species 2 in
Eq.~(\ref{particleCons}) is half that of the original dimer from
Eq.~(\ref{particleCons_nonlin}): $\mu_2/2 \to \mu_2$.  Above said, we
will continue to call species one and two ``the monomer'' and ``the
dimer,'' respectively.

\begin{figure}[t]
\centering
\includegraphics[width= \figurewidth]{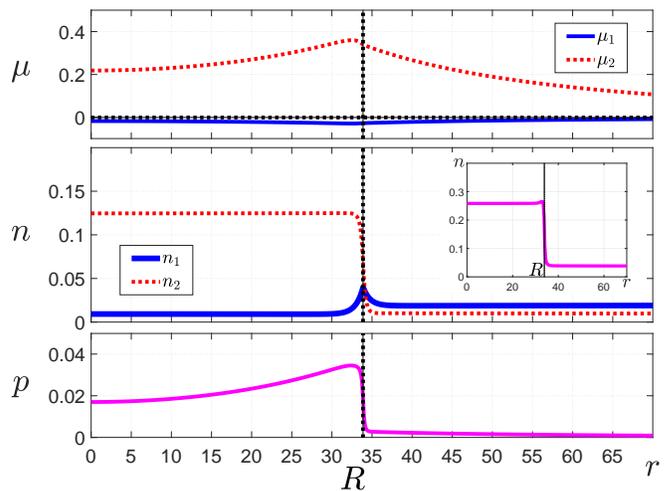}
\caption{\label{F:munpr_stat} First-order reaction case from
  Eq.~(\ref{particleCons}): The radial-coordinate dependences of the
  chemical potential $\mu_i$, concentrations $n_i$, and pressure $p$
  for a stationary, spherically symmetric cluster. The inset show the
  $r$-dependence of the total amount of the solute, $n \equiv n_1 + 2
  n_2$. The following parameter values are employed:
  $\kappa^\text{(d)}_1 = \kappa^\text{(d)}_2 = \kappa^\text{(m)}_1 =
  \kappa^\text{(m)}_2 = 40$, $m^\text{(d)}_1 = m^\text{(m)}_1 =
  52.36$, $m^\text{(d)}_2 = m^\text{(m)}_2 = 500$, $n^\text{(d)}_1 =
  0.01$, $n^\text{(d)}_2 = 0.12$, $n^\text{(m)}_1 = 0.02$,
  $n^\text{(m)}_2 = 0.01$, $D^\text{(d)}_1 = 0.33$, $D^\text{(d)}_2 =
  0.25$, $D^\text{(m)}_1 = 1$, $D^\text{(m)}_2 = 0.76$,
  $k^\text{(d)}_1 = 0.001$, $k^\text{(d)}_2 = 0.000077$,
  $k^\text{(m)}_1 = 0.000038$, $k^\text{(m)}_2 = 0.000077$, $\Delta g
  = 0.01$, $k_1^\ddag = 0.00005$, $k_2^\ddag = 0.00003$. The units are
  arbitrary; the unit of length can be thought of as roughly
  comparable to molecular dimensions and the unit of energy to
  $k_BT$. The values for the rate coefficients and diffusivities were
  chosen to yield values for the cluster size comparable to those seen
  in protein solutions.}
\end{figure}

Eqs.~(\ref{particleCons}) can also be considered on their own merit.
They can approximate a physical situation where the species 1 converts
into species 2 by binding a third species that is part of the
buffer. If the transport of this third species is fast compared with
the transport of species 1 and 2, then the above equations apply.
Mathematically, the linearity of Eqs.~(\ref{particleCons}) renders the
problem linear in respective pure phases and thus reduces the
differential equation to an algebraic characteristic equation than can
be solved much more readily than the original non-linear differential
equation. This allows one to readily explore broad ranges of
parameters. Once a non-trivial solution of the 1st order case
(\ref{particleCons}) is found, one may then attempt to confirm whether
a similar solution exists in the more complicated, 2nd order case from
Eqs.~(\ref{particleCons_nonlin}). Throughout, we consider exclusively
the spherically symmetric geometry; such solutions are expected to
minimize the surface tension between the two phases during phase
coexistence.~\cite{CL_LG}

\begin{figure}[t]
\centering
\includegraphics[width= .9 \figurewidth]{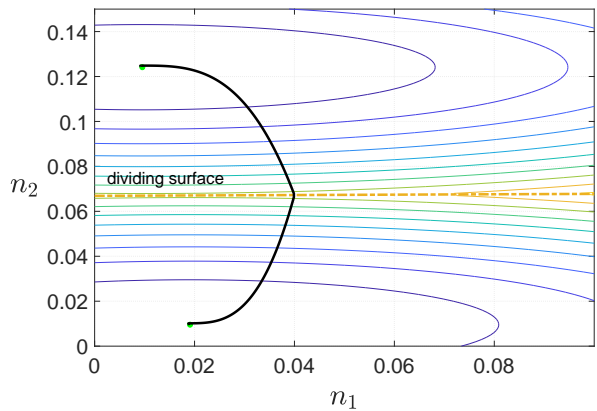}
\caption{\label{F:contour} Contour plot of the bulk free energy
  density $\cV(n_1, n_2)$ from Eq.~(\ref{ftilde2}) as a function of
  the concentrations $n_1$ and $n_2$ of the components.  The two
  paraboloids, corresponding with the free energies of the two phases,
  intersect at the ``dividing surface.'' The upper-left minimum
  corresponds to the dimer-rich solution, which is the minority
  phase. The curve connecting the two minima is the parametric plot of
  the concentrations $n_1$ and $n_2$ from Fig.~\ref{F:munpr_stat}, the
  parameter being the radial coordinate $r$.}
\end{figure}

We specifically inquire whether long-lived inclusions of the
dimer-rich phase could form inside the monomer-rich phase, when the
dimer-rich phase is in fact metastable:
\begin{equation} \label{Deltag} \Delta g \equiv g^\text{(d)} -
  g^\text{(m)} > 0.
\end{equation}
(When $\Delta g < 0$, nucleation would proceed even in the absence of
monomer-dimer conversion, of course. This case, albeit of some
interest, will not be considered here.) Such long-lived inclusions, if
any, could represent a metastable state and/or nucleate in an
activated fashion.  In either case, we must look for {\em stationary}
droplet-like solutions for Eqs.~(\ref{Fick})-(\ref{particleCons}): $
\dot{n}_i = 0$, where the minority and majority phase are the
dimer-rich and monomer-rich liquids, respectively.

Such non-trivial stationary solutions do indeed exist as we exemplify
in Fig.~\ref{F:munpr_stat}. There we show the coordinate dependences
of the chemical potentials and concentrations of the two species, and
the hydrostatic pressure. (The coordinate-dependent pressure was
computed as in Ref.~\onlinecite{CL_LG}, see Methods.) A parametric
plot of the concentrations of the monomer and dimer, the parameter
being the distance from the droplet center, is shown in
Fig.~\ref{F:contour}. There we also show the contour plot of the bulk
free energy $\cV$ from Eq.~(\ref{ftilde2}).  The length $R$ denotes
the radius of the spherical region occupied by the dimer-rich
phase. The value of $R$ is determined self-consistently as a result of
solving the equations. We will use $R$ as the nominal cluster radius
but note that it is a lower bound on the cluster size because the
concentrations reach their bulk values at $r > R$, as should be clear
from Fig.~\ref{F:munpr_stat}.

As anticipated by Pan et al.~\cite{PVL}, the chemical potentials in
Fig.~\ref{F:munpr_stat} tend exponentially rapidly to their bulk
values at large separations from the droplet center, in steady
state. Likewise, the net particle exchange for each individual species,
between the droplet and the bulk solution, drops exponentially fast
into the bulk.  Nearer to the droplet, there is significant influx of
the monomer toward the droplet and outflaw of the dimer, accompanied
by a net decay of the dimer into the monomer. At the same time, the
total flux of the solute, i.e., the quantity $\sum_i \widetilde D_i
\nabla \mu_i$, is identically zero in steady state.

The situation {\em inside} the droplet is drastically different from
that anticipated in Ref.~\onlinecite{PVL} in that it largely mirrors
the transport pattern on the outside: For the most part, the monomer
flows from the center to the boundary while the dimer does the
opposite. Figs.~\ref{F:munpr_stat} and \ref{F:contour} highlight a
peculiar nature of the stationary solution at $\Delta g > 0$: Both
chemical potentials, the concentration of the monomer, and the
pressure exhibits non-monotonic dependences on the radial coordinate
$r$. In contrast, such dependences are expected to be monotonic during
conventional nucleation.~\cite{CL_LG} (Furthermore, the chemical
potentials are strictly spatially uniform when the droplet is
critical~\cite{Bray, CL_LG}!)  We show separately the quantity $n_m
\equiv n_1 + 2 n_2$, which is the total concentration of the solute,
irrespective of whether it is in the form of monomer or dimer.
According to Fig.~\ref{F:munpr_stat}, there a small pile up of the
solute at the droplet boundary.

The apparent decrease in the pressure toward the center of the droplet
is expected because the pressure difference between the bulk
dimer-rich and monomer-rich phases is the negative of the bulk free
energy difference:~\cite{CL_LG}
\begin{equation} \label{Deltap} p_\text{bulk}^\text{(d)} -
  p_\text{bulk}^\text{(m)} = - \Delta g.
\end{equation}

\begin{figure}[t]
\centering
\includegraphics[width=.9\figurewidth]{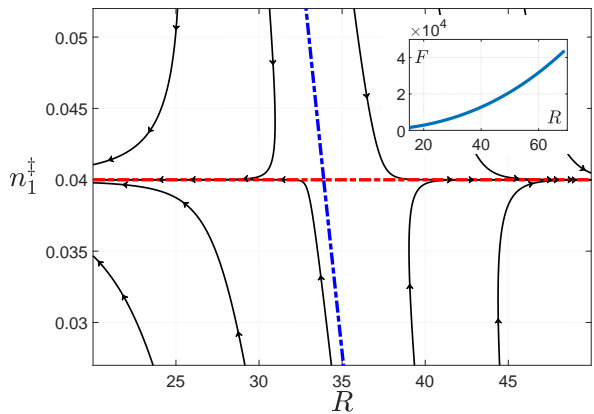}
\caption{\label{F:flow} The flow chart for the clusters size $R$ and
  composition $n^\ddag_1$ of the monomer at the boundary. The
  stationary solution is at the intersection of the blue and red
  dashed lines. The parameter values are the same as in
  Fig.~\ref{F:munpr_stat}.}
\end{figure}

We next ask whether the above droplet solution is a metastable state
or a transition state configuration. To answer this question, we do
not attempt to solve the full-blown time dependent problem.  Instead,
we first artificially constrain the values of the concentrations at
droplet boundary and the droplet radius $R$ away from their stationary
values. We then use the resulting profiles of the chemical potentials
to determine the fluxes of the monomer and dimer at the cluster
boundary. In turn, these fluxes are used to estimate the value of the
time derivatives $\dot{R}$ and $\dot{n}_1^\ddagger$, where
$n_1^\ddagger$ is the concentration of the monomer at the
boundary. ($n_2^\ddagger$ is specified automatically because the
boundary is a line in the $(n_1^\ddagger, n_2^\ddagger)$ plane.)
Finally, we make a flow chart corresponding to the vector $(\dot{R},
\dot{n}_1^\ddagger)$ in the $(R, n_1^\ddagger)$ plane, as shown in
Fig.~\ref{F:flow}. This flow chart demonstrates that the stationary
solution is, in fact, a critical point beyond which the droplet will
grow indefinitely but evaporate otherwise. At the same time, we note
the free energy of the droplet is a monotonically increasing function
of the droplet radius, as we show in the inset of Fig.~\ref{F:flow}.
When combined, these two notions would seem to cast serious doubts on
the present analysis because they indicate the droplet will grow
indefinitely despite its free energy increasing in the process.  This
would seem to contradict the second law of thermodynamics.

\begin{figure}[t]
\centering
\includegraphics[width= .7 \figurewidth]{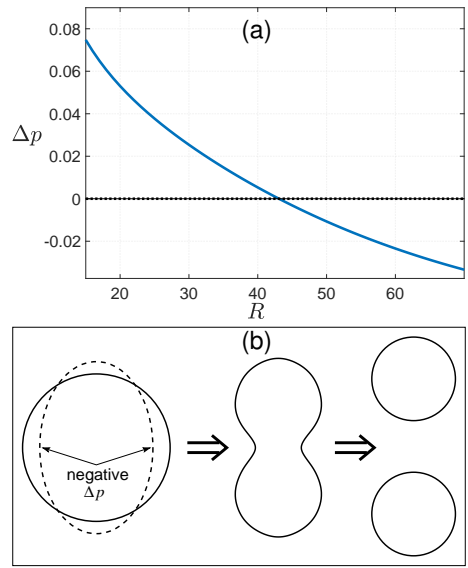}
\caption{\label{pvsr} \textbf{(a)} The pressure differential between
  the cluster centre and the solution bulk, as a function of the
  droplet radius. Only one point corresponds to a stationary solution,
  which is the same as that in Fig.~\ref{F:flow}. \textbf{(b)} A
  graphical explanation of the mechanical instability and subsequent
  breaking of a droplet as the pressure differential becomes
  negative.}
\end{figure}

To resolve this seeming contradiction, we again employ
Eq.~(\ref{Deltap}). For sufficiently large droplets the pressure
inside, relative to outside, will be negative implying the droplet
will be {\em mechanically} unstable.  This can be directly seen in
Fig.~\ref{pvsr}(a), where we plot the pressure $\Delta p$ in the
center of the sphere, relative to its value in the solution bulk, as a
function of the droplet radius $R$ for a spherical droplet. We observe
that beyond a certain threshold value of $R$, $\Delta p$ becomes
negative. (We exemplify such a solution in Methods.)  This threshold
value of $R$ is a lower bound on the cluster size beyond which the
cluster would become mechanically unstable. To see this, we first note
that because of shape fluctuations, some parts of the boundary have
lower curvature than $2/R$. Once the pressure under the boundary
becomes sufficiently low on average, the inner regions adjacent to the
flatter parts of the boundary will be at a negative pressure, relative
to $p^\text{(m)}_\text{bulk}$. This is illustrated in
Fig.~\ref{pvsr}(b). Consequently, the boundary will begin to cave in,
around these areas of lowered pressure; this will lead to further
lowering of the pressure around those areas and, eventually, a
breakdown via catastrophic caving or necking.

The following microscopic picture thus emerges: In a steady-state
solution, clusters continuously nucleate, grow, and ultimately decay
because of a mechanical instability. The latter instability ultimately
stems from the dimer-rich phase being thermodynamically metastable. No
problems with the second law arise for an {\em ensemble} of clusters:
For each nucleating cluster, there is a decaying one, in steady-state,
and so there is no net entropy production or consumption. The
monotonic increase of the free energy of an individual droplet with
the droplet size drives home the notion that the clusters are
stabilized {\em kinetically}, not thermodynamically.  The
stabilization comes about because once formed, as a result of density
fluctuations, a dimer-rich region will extend for distances dimers
will travel before they decay back into monomers.

\begin{figure}[t]
\centering
\includegraphics[width= .8 \linewidth]{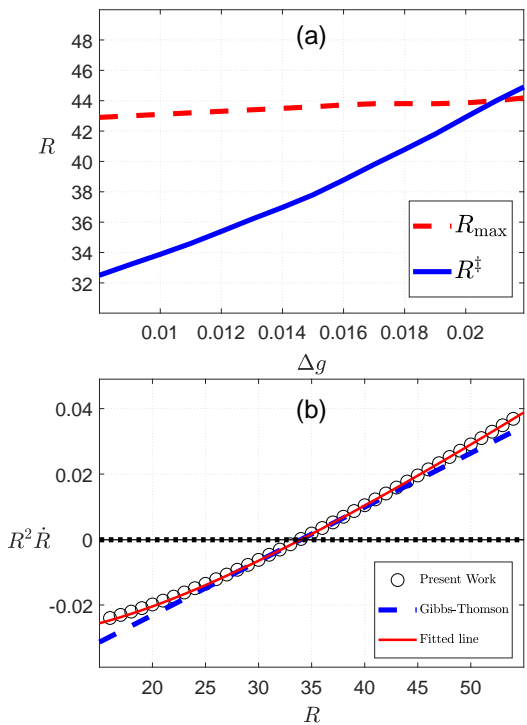}
\caption{\label{F:Fig5} \textbf{(a)} The $\Delta g$ dependences of the
  critical radius $R^\ddagger$ and the threshold value of the cluster
  $R_\text{max}$ beyond which the cluster becomes mechanically
  unstable. \textbf{(b)} The dependence of the rate of volumetric
  growth of the cluster as a function of its radius. Only one point on
  this curve corresponds to a stationary solution, as in
  Fig.~\ref{pvsr}(a).}
\end{figure}

One may further elaborate on the above notions of kinetic
stabilization. The reaction terms in Eqs.~(\ref{particleCons_nonlin})
and (\ref{particleCons}) are local and thus the kinetic stabilization,
if any, would be of bulk character. On the other hand, such
stabilization can operate only on lengths not exceeding the kinetic
lengths of the type in Eq.~(\ref{RDk}). Thus we conclude that for the
present scenario to be viable, the parameter values should be such
that the critical size $R^\ddagger$ for nucleation should be less than
the pertinent kinetic length. We can check this notion, even if
somewhat indirectly, by computing the critical size for a range of
$\Delta g$ values. Larger values of $\Delta g$ should imply less
overall stabilization---thermodynamic plus kinetic---and,
consequently, larger values for the critical radius. This is borne out
by the results in Fig.~\ref{F:Fig5}(a). In that Figure, we also show
the dependence of the threshold value of the droplet radius
$R_\text{max}$, at which the pressure differential $\Delta p$ in the
center of the droplet would vanish. We observe that, indeed, there is
an upper limit on the bulk free energy excess of the dimer-rich phase
beyond which already sub-critical clusters would be mechanically
unstable and, thus, could not emerge in the first place.  Because the
characteristic equations are complicated, it is difficult to see the
explicit dependence of the lengths in Fig.~\ref{F:Fig5}(a) on the
kinetic coefficients. We have checked that for specific values of
parameters, the critical radius $R^\ddagger$ does decrease with the
decay rate $k_2$ of the dimer, consistent with the heuristic arguments
of Pan et al.~\cite{PVL}; the corresponding data can be found in
Methods.

Fig.~\ref{F:Fig5}(a) indicates the range of possible values for the
cluster size. Indeed, because sub-critical clusters would rapidly
evaporate, one should readily observe clusters only within the size
range specified by the critical size $R^\ddagger$ and the threshold
size $R_\text{max}$ (if $R^\ddagger < R_\text{max}$). According to
Fig.~\ref{F:Fig5}(a), this range is relatively modest, consistent with
the apparent weak dependence of the cluster size on the concentration
of the solute.

Next, we address the question of the ripening of the clusters in a
freshly prepared solution. According to
Refs.~\onlinecite{LifshitzSlyozov1961, WagnerOR, Bray,
  PhysRevA.20.595}, conventional Ostwald ripening comes about for the
following reason: If the typical size of the minority phase is
sub-macroscopic, the solution of the pertinent species in the majority
phase is over-saturated, the degree of supersaturation decreasing with
the typical droplet size according to the usual Gibbs-Thompson
relation.~\cite{LLstat} At a given value of supersaturation, droplets
smaller than the corresponding critical size will evaporate, while
droplets that are bigger than the critical size will grow. As a
result, the average droplet size will grow until the minority phase
reaches macroscopic dimensions while the supersaturation peters out.
Within the Gibbs-Thompson approximation and in the limit of diffusion
controlled droplet growth, the volumetric rate of droplet growth, $R^2
\dot{R}$ happens to scale linearly with the deviation $(R-R^\ddagger)$
of the droplet radius from the critical radius, see Methods. This is
shown by the dashed line in Fig.~\ref{F:Fig5}(b).  The corresponding
dependence for the present clusters is shown that Figure with
symbols. Although different from a strict linear form, the quantity
$R^2 \dot{R}$ for {\em kinetically} stabilized clusters is still a
monotonically increasing function of $R$ vanishing at
$R=R^\ddagger$. As detailed in Methods, the data in
Fig.~\ref{F:Fig5}(b) imply that well before the steady-state cluster
is reached, clusters grow according to a power-law $R^\ddag \propto =
t^{0.32 \pm 0.01}$. This is quite close to if not somewhat faster than
the dependence $t^{0.26 \pm 0.03}$ observed by Ye Li et
al.~\cite{LLVFPostwaldRipening}. For comparison, the Lifshitz-Slyozov
mechanism of conventional Oswald ripening predicts $R^\ddag \propto =
t^{1/3}$.

\begin{figure}[t]
\centering
\includegraphics[width= \linewidth]{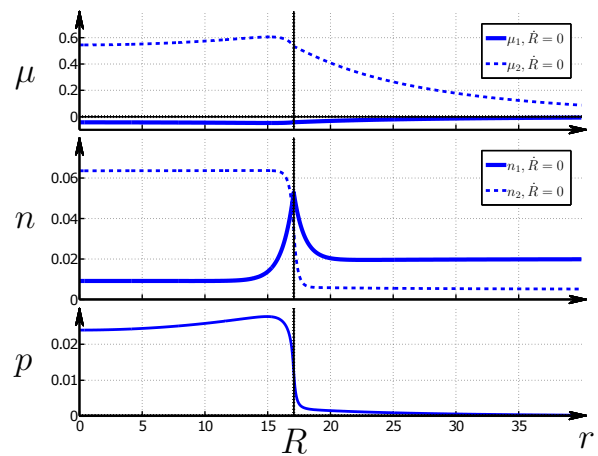}
\caption{\label{F:munpr_stat_nonlin} Second-order reaction case from
  Eq.~(\ref{particleCons_nonlin}): The radial-coordinate dependences
  of the chemical potential $\mu_i$, concentrations $n_i$, and
  pressure $p$ for a stationary, spherically symmetric cluster.  The
  following parameter values are employed: $\kappa^\text{(d)}_1 =
  \kappa^\text{(d)}_2 = \kappa^\text{(m)}_1 = \kappa^\text{(m)}_2 =
  40$, $m^\text{(d)}_1 = m^\text{(m)}_1 = 52.36$, $m^\text{(d)}_2 =
  m^\text{(m)}_2 = 500$, $n^\text{(d)}_1 = 0.01$, $n^\text{(d)}_2 =
  0.063$, $n^\text{(m)}_1 = 0.02$, $n^\text{(m)}_2 = 0.005$,
  $D^\text{(d)}_1 = 0.67$, $D^\text{(d)}_2 = 0.25$, $D^\text{(m)}_1 =
  2$, $D^\text{(m)}_2 = 0.76$, $k^\text{(d)}_1 = 0.1$, $k^\text{(d)}_2
  = 0.00008$, $k^\text{(m)}_1 = 0.025$, $k^\text{(m)}_2 = 0.001$,
  $\Delta g = 0.01$.}
\end{figure}

This notion suggests that an Ostwald-like ripening could take place in
cluster-containing solutions. Indeed, according to
Fig.~\ref{F:Fig5}(a), the critical radius increases with $\Delta g$,
as already mentioned. On the other hand, $\Delta g$ increases with
lowering of the concentration of the solute in the bulk solution
($g^\text{(m)} \sim \ln n_{1, b}$).  In a freshly prepared solution,
the typical cluster size is less than its value in equilibrium
resulting in an excess solute to compensate for the excess curvature of
the cluster surface. As the average cluster size increases, the amount
of this excess solute will decrease leading to an increase in $\Delta
g$, and, consequently an increase in the critical radius. The increase
of the critical radius with time is similar to what happens during
conventional Ostwald ripening. In contrast with the conventional
Ostwald ripening, however, the supersaturation due to the finite
curvature {\em increases}, not decreases with time. This is because
the minority phase here is thermodynamically metastable in the first
place. Yet as in the case of growth of an individual droplet, the
seemingly ``positive-feedback loop'' for $\Delta g$ does not lead to a
runaway growth of the droplets because of the mechanical instability
discussed earlier.  Furthermore, as $\Delta g$ approaches its limiting
value, where $R^\ddagger = R_\text{max}$, the time dependence of the
typical cluster size must stop following the Ostwald-like $t^{1/3}$
and, instead, level off at the equilibrium value of $R_\text{max}$.

Finally, we present in Fig.~\ref{F:munpr_stat_nonlin} the stationary
solution corresponding to the original second-order reaction setup
from Eqs.~(\ref{particleCons_nonlin}). This solution was obtained
using the finite element method~\cite{ascher1988numerical} and
requires much more effort than the first-order case, both in terms of
implementation and computation proper; see Methods for details. In any
event, we observe that the non-linearity in the reaction terms does
not destroy kinetic stabilization observed in the case of first order
kinetics; the two cases produces qualitatively similar results. At the
same time, we note introducing the non-linearity in the reaction
kinetics has very substantial quantitative effects. For instance, for
the same values of the parameters that yield a droplet solution when
the complexation reaction is second order, the corresponding
linearized case may not exhibit a droplet solution altogether,
steady-state or not.

{\bf Acknowledgments:} We thank Peter G. Vekilov and Peter G. Wolynes
for many inspiring conversations.  We gratefully acknowledge the
support by the NSF Grant MCB-1244568 and the Welch Foundation Grant
No. E-1765.

\bibliographystyle{naturemag}
\bibliography{lowT}

\begin{thebibliography}{10}
\expandafter\ifx\csname url\endcsname\relax
  \def\url#1{\texttt{#1}}\fi
\expandafter\ifx\csname urlprefix\endcsname\relax\def\urlprefix{URL }\fi
\providecommand{\bibinfo}[2]{#2}
\providecommand{\eprint}[2][]{\url{#2}}

\bibitem{Elbaum-Garfinkle7189}
\bibinfo{author}{Elbaum-Garfinkle, S.} \emph{et~al.}
\newblock \bibinfo{title}{{The disordered P granule protein LAF-1 drives phase
  separation into droplets with tunable viscosity and dynamics}}.
\newblock \emph{\bibinfo{journal}{Proc. Natl. Acad. Sci. U.~S.~A.}}
  \textbf{\bibinfo{volume}{112}}, \bibinfo{pages}{7189--7194}
  (\bibinfo{year}{2015}).

\bibitem{BanjadeE6426}
\bibinfo{author}{Banjade, S.} \emph{et~al.}
\newblock \bibinfo{title}{{Conserved interdomain linker promotes phase
  separation of the multivalent adaptor protein Nck}}.
\newblock \emph{\bibinfo{journal}{Proc. Natl. Acad. Sci. U.~S.~A.}}
  \textbf{\bibinfo{volume}{112}}, \bibinfo{pages}{E6426--E6435}
  (\bibinfo{year}{2015}).

\bibitem{Shineaaf4382}
\bibinfo{author}{Shin, Y.} \& \bibinfo{author}{Brangwynne, C.~P.}
\newblock \bibinfo{title}{Liquid phase condensation in cell physiology and
  disease}.
\newblock \emph{\bibinfo{journal}{Science}} \textbf{\bibinfo{volume}{357}}
  (\bibinfo{year}{2017}).

\bibitem{0034-4885-81-4-046601}
\bibinfo{author}{Berry, J.}, \bibinfo{author}{Brangwynne, C.~P.} \&
  \bibinfo{author}{Haataja, M.}
\newblock \bibinfo{title}{Physical principles of intracellular organization via
  active and passive phase transitions}.
\newblock \emph{\bibinfo{journal}{Rep. Progress Phys.}}
  \textbf{\bibinfo{volume}{81}}, \bibinfo{pages}{046601}
  (\bibinfo{year}{2018}).

\bibitem{GibbsV1}
\bibinfo{author}{Gibbs, J.~W.}
\newblock \emph{\bibinfo{title}{The Scientific Papers of J. Willard Gibbs, Vol.
  1: Thermodynamics}} (\bibinfo{publisher}{Ox Bow Press},
  \bibinfo{address}{Connecticut}, \bibinfo{year}{1993}).

\bibitem{LLstat}
\bibinfo{author}{Landau, L.~D.} \& \bibinfo{author}{Lifshitz, E.~M.}
\newblock \emph{\bibinfo{title}{Statistical Mechanics}}
  (\bibinfo{publisher}{Pergamon Press}, \bibinfo{address}{New York},
  \bibinfo{year}{1980}).

\bibitem{GlikoJACS2005}
\bibinfo{author}{Gliko, O.} \emph{et~al.}
\newblock \bibinfo{title}{A metastable prerequisite for the growth of lumazine
  synthase crystals}.
\newblock \emph{\bibinfo{journal}{JACS}} \textbf{\bibinfo{volume}{127}},
  \bibinfo{pages}{3433} (\bibinfo{year}{2005}).

\bibitem{Georgalis1999}
\bibinfo{author}{Georgalis, Y.}, \bibinfo{author}{Umbach, P.},
  \bibinfo{author}{Saenger, W.}, \bibinfo{author}{Ihmels, B.} \&
  \bibinfo{author}{Soumpasis, D.~M.}
\newblock \bibinfo{title}{Ordering of fractal clusters in crystallizing
  lysozyme solutions}.
\newblock \emph{\bibinfo{journal}{J. Amer. Chem. Soc.}}
  \textbf{\bibinfo{volume}{121}}, \bibinfo{pages}{1627--1635}
  (\bibinfo{year}{1999}).

\bibitem{doi:10.1021/jp068827o}
\bibinfo{author}{Gliko, O.} \emph{et~al.}
\newblock \bibinfo{title}{Metastable liquid clusters in super- and
  undersaturated protein solutions}.
\newblock \emph{\bibinfo{journal}{J. Phys. Chem. B}}
  \textbf{\bibinfo{volume}{111}}, \bibinfo{pages}{3106--3114}
  (\bibinfo{year}{2007}).

\bibitem{PVL}
\bibinfo{author}{Pan, W.}, \bibinfo{author}{Vekilov, P.~G.} \&
  \bibinfo{author}{Lubchenko, V.}
\newblock \bibinfo{title}{Origin of anomalous mesoscopic phases in protein
  solutions}.
\newblock \emph{\bibinfo{journal}{J. Phys. Chem. B}}
  \textbf{\bibinfo{volume}{114}}, \bibinfo{pages}{7620--7630}
  (\bibinfo{year}{2010}).

\bibitem{LLVFPostwaldRipening}
\bibinfo{author}{Li, Y.}, \bibinfo{author}{Lubchenko, V.},
  \bibinfo{author}{Vorontsova, M.~A.}, \bibinfo{author}{Filobelo, L.} \&
  \bibinfo{author}{Vekilov, P.~G.}
\newblock \bibinfo{title}{Ostwald-like ripening of the anomalous mesoscopic
  clusters in protein solutions}.
\newblock \emph{\bibinfo{journal}{J. Phys. Chem. B}}
  \textbf{\bibinfo{volume}{116}}, \bibinfo{pages}{10657--10664}
  (\bibinfo{year}{2012}).

\bibitem{YEARLEY20141763}
\bibinfo{author}{Yearley, E.~J.} \emph{et~al.}
\newblock \bibinfo{title}{Observation of small cluster formation in
  concentrated monoclonal antibody solutions and its implications to solution
  viscosity}.
\newblock \emph{\bibinfo{journal}{Biophys. J.}} \textbf{\bibinfo{volume}{106}},
  \bibinfo{pages}{1763--1770} (\bibinfo{year}{2014}).

\bibitem{SleutelE546}
\bibinfo{author}{Sleutel, M.} \& \bibinfo{author}{Van~Driessche, A. E.~S.}
\newblock \bibinfo{title}{Role of clusters in nonclassical nucleation and
  growth of protein crystals}.
\newblock \emph{\bibinfo{journal}{Proc. Natl. Acad. Sci. U.~S.~A.}}
  \textbf{\bibinfo{volume}{111}}, \bibinfo{pages}{E546--E553}
  (\bibinfo{year}{2014}).

\bibitem{doi:10.1021/acs.cgd.6b01826}
\bibinfo{author}{Schubert, R.} \emph{et~al.}
\newblock \bibinfo{title}{Real-time observation of protein dense liquid cluster
  evolution during nucleation in protein crystallization}.
\newblock \emph{\bibinfo{journal}{Crystal Growth \& Design}}
  \textbf{\bibinfo{volume}{17}}, \bibinfo{pages}{954--958}
  (\bibinfo{year}{2017}).

\bibitem{Gliko2005}
\bibinfo{author}{Gliko, O.} \emph{et~al.}
\newblock \bibinfo{title}{Dense liquid droplets as a step source for the
  crystallization of lumazine synthase}.
\newblock \emph{\bibinfo{journal}{Journal of Cryst. Growth}}
  \textbf{\bibinfo{volume}{275}}, \bibinfo{pages}{e1409}
  (\bibinfo{year}{2005}).

\bibitem{doi:10.1021/acs.cgd.7b01299}
\bibinfo{author}{Warzecha, M.}, \bibinfo{author}{Safari, M.~S.},
  \bibinfo{author}{Florence, A.~J.} \& \bibinfo{author}{Vekilov, P.~G.}
\newblock \bibinfo{title}{Mesoscopic solute-rich clusters in olanzapine
  solutions}.
\newblock \emph{\bibinfo{journal}{Crystal Growth \& Design}}
  \textbf{\bibinfo{volume}{17}}, \bibinfo{pages}{6668--6676}
  (\bibinfo{year}{2017}).

\bibitem{UZUNOVA20101976}
\bibinfo{author}{Uzunova, V.~V.}, \bibinfo{author}{Pan, W.},
  \bibinfo{author}{Galkin, O.} \& \bibinfo{author}{Vekilov, P.~G.}
\newblock \bibinfo{title}{Free heme and the polymerization of sickle cell
  hemoglobin}.
\newblock \emph{\bibinfo{journal}{Biophys. J.}} \textbf{\bibinfo{volume}{99}},
  \bibinfo{pages}{1976 -- 1985} (\bibinfo{year}{2010}).

\bibitem{C2FD20058A}
\bibinfo{author}{Uzunova, V.}, \bibinfo{author}{Pan, W.},
  \bibinfo{author}{Lubchenko, V.} \& \bibinfo{author}{Vekilov, P.~G.}
\newblock \bibinfo{title}{Control of the nucleation of sickle cell hemoglobin
  polymers by free hematin}.
\newblock \emph{\bibinfo{journal}{Faraday Discuss.}}
  \textbf{\bibinfo{volume}{159}}, \bibinfo{pages}{87--104}
  (\bibinfo{year}{2012}).

\bibitem{Yamazaki2154}
\bibinfo{author}{Yamazaki, T.} \emph{et~al.}
\newblock \bibinfo{title}{Two types of amorphous protein particles facilitate
  crystal nucleation}.
\newblock \emph{\bibinfo{journal}{Proc. Natl. Acad. Sci. U.~S.~A.}}
  \textbf{\bibinfo{volume}{114}}, \bibinfo{pages}{2154--2159}
  (\bibinfo{year}{2017}).

\bibitem{doi:10.1063/1.1943389}
\bibinfo{author}{Kashchiev, D.}, \bibinfo{author}{Vekilov, P.~G.} \&
  \bibinfo{author}{Kolomeisky, A.~B.}
\newblock \bibinfo{title}{Kinetics of two-step nucleation of crystals}.
\newblock \emph{\bibinfo{journal}{J. Chem. Phys.}}
  \textbf{\bibinfo{volume}{122}}, \bibinfo{pages}{244706}
  (\bibinfo{year}{2005}).

\bibitem{LifshitzSlyozov1961}
\bibinfo{author}{Lifshitz, I.~M.} \& \bibinfo{author}{Slyozov, V.~V.}
\newblock \bibinfo{title}{The kinetics of precipitation from supersaturated
  solid solutions}.
\newblock \emph{\bibinfo{journal}{J. Phys. Chem. Solids}}
  \textbf{\bibinfo{volume}{19}}, \bibinfo{pages}{35--50}
  (\bibinfo{year}{1961}).

\bibitem{WagnerOR}
\bibinfo{author}{Wagner, C.}
\newblock \bibinfo{title}{Theorie der alterung von niederschlägen durch
  umlösen (ostwald-reifung)}.
\newblock \emph{\bibinfo{journal}{Elektrochem.}} \textbf{\bibinfo{volume}{65}},
  \bibinfo{pages}{581--591} (\bibinfo{year}{1961}).

\bibitem{Bray}
\bibinfo{author}{Bray, A.~J.}
\newblock \bibinfo{title}{Theory of phase-ordering kinetics}.
\newblock \emph{\bibinfo{journal}{Adv. Phys.}} \textbf{\bibinfo{volume}{43}},
  \bibinfo{pages}{357--459} (\bibinfo{year}{1994}).

\bibitem{PhysRevA.20.595}
\bibinfo{author}{Siggia, E.~D.}
\newblock \bibinfo{title}{Late stages of spinodal decomposition in binary
  mixtures}.
\newblock \emph{\bibinfo{journal}{Phys. Rev. A}} \textbf{\bibinfo{volume}{20}},
  \bibinfo{pages}{595--605} (\bibinfo{year}{1979}).

\bibitem{C5SM02234G}
\bibinfo{author}{Lutsko, J.~F.} \& \bibinfo{author}{Nicolis, G.}
\newblock \bibinfo{title}{Mechanism for the stabilization of protein clusters
  above the solubility curve}.
\newblock \emph{\bibinfo{journal}{Soft Matter}} \textbf{\bibinfo{volume}{12}},
  \bibinfo{pages}{93--98} (\bibinfo{year}{2016}).

\bibitem{0953-8984-28-24-244020}
\bibinfo{author}{Lutsko, J.~F.}
\newblock \bibinfo{title}{Mechanism for the stabilization of protein clusters
  above the solubility curve: the role of non-ideal chemical reactions}.
\newblock \emph{\bibinfo{journal}{J. Phys. Cond. Matter}}
  \textbf{\bibinfo{volume}{28}}, \bibinfo{pages}{244020}
  (\bibinfo{year}{2016}).

\bibitem{Goldenfeld}
\bibinfo{author}{Goldenfeld, N.}
\newblock \emph{\bibinfo{title}{Lectures on phase transitions and the
  renormalization group}} (\bibinfo{publisher}{Addison-Wesley},
  \bibinfo{address}{Reading, MA}, \bibinfo{year}{1992}).

\bibitem{L_AP}
\bibinfo{author}{Lubchenko, V.}
\newblock \bibinfo{title}{Theory of the structural glass transition: A
  pedagogical review}.
\newblock \emph{\bibinfo{journal}{Adv. Phys.}} \textbf{\bibinfo{volume}{64}},
  \bibinfo{pages}{283--443} (\bibinfo{year}{2015}).

\bibitem{PhysRev.137.A1441}
\bibinfo{author}{Mermin, N.~D.}
\newblock \bibinfo{title}{Thermal properties of the inhomogeneous electron
  gas}.
\newblock \emph{\bibinfo{journal}{Phys. Rev.}} \textbf{\bibinfo{volume}{137}},
  \bibinfo{pages}{A1441--A1443} (\bibinfo{year}{1965}).

\bibitem{CahnHilliard}
\bibinfo{author}{Cahn, H.~W.} \& \bibinfo{author}{Hilliard, J.~E.}
\newblock \bibinfo{title}{{Free Energy of Nonuniform System. I. Interfacial
  Free Energy}}.
\newblock \emph{\bibinfo{journal}{J. Chem. Phys.}}
  \textbf{\bibinfo{volume}{28}}, \bibinfo{pages}{258--267}
  (\bibinfo{year}{1958}).

\bibitem{CL_LG}
\bibinfo{author}{Chan, H.~Y.} \& \bibinfo{author}{Lubchenko, V.}
\newblock \bibinfo{title}{{Pressure in the Landau-Ginzburg functional: Pascal's
  law, nucleation in fluid mixtures, a meanfield theory of amphiphilic action,
  and interface wetting in glassy liquids}}.
\newblock \emph{\bibinfo{journal}{J. Chem. Phys.}}
  \textbf{\bibinfo{volume}{143}}, \bibinfo{pages}{124502}
  (\bibinfo{year}{2015}).

\bibitem{ascher1988numerical}
\bibinfo{author}{Ascher, U.}, \bibinfo{author}{Mattheij, R.} \&
  \bibinfo{author}{Russell, R.}
\newblock \emph{\bibinfo{title}{Numerical Solution of Boundary Value Problems
  for Ordinary Differential Equations}}.
\newblock Classics in Applied Mathematics (\bibinfo{publisher}{Society for
  Industrial and Applied Mathematics}, \bibinfo{year}{1988}).

\end{thebibliography}

\newpage

\begin{widetext}
\end{widetext}

\newpage

\section*{Methods}

To describe phase coexistence we employ a double-minimum bulk free
energy density $\cV\left(n_1, n_2\right)$.  The latter free energy
corresponds with the grand-canonical ensemble and is straightforwardly
related to the Helmholtz free energy density $f$:~\cite{CL_LG}
\begin{equation}\label{ftilde}
  \cV \left(n_1, n_2\right) = f\left(n_1, n_2\right) - \mu_{1,
    b} n_1 - \mu_{2, b} n_2.
\end{equation}
where $\mu_{i, b}$ is the chemical potential of species $i$ in the
bulk.

A smooth surface exhibiting two minima has to be a quartic polynomial
or a more complicated function, which renders even the otherwise
linear differential equations (\ref{particleCons}) highly non-linear
and difficult to solve even numerically. To circumvent this
difficulty, we employ a bulk free energy which is not smooth but,
instead, consists of two intersecting paraboloids, see
Eq.~(\ref{ftilde2}) and Fig.~\ref{F:contour}. The resulting free
energy surface exhibits a singularity, in the form of a discontinuous
gradient, where the two paraboloids from Eq.~(\ref{ftilde2}), $\alpha
= \text{m}$ and $\alpha = \text{d}$, intersect.  The singularity is
however confined to a region of measure zero, the latter region
corresponding to the phase boundary. In each individual phase, the
transport part of the problem reduces to linear differential
equations. The respective solutions must be patched together where the
bulk free energy is singular, i.e., at the phase boundary.  Patching
such solutions for mixtures, as opposed to systems described by only
one order parameter, presents some subtlety and has been worked out
earlier by us.~\cite{CL_LG}
\begin{align}
  &n_i(R^+) = n_i(R^-) \equiv n_i^\ddag  \label{eq:BC1}\\
  &\mu_i(R^-) = \mu_i(R^+) \label{eq:BC4}\\
  &\sum_i \left. \kappa_i (\prtl n_i/\prtl r)^2 \right|_{R^-}^{R^+} =
  0.
  \label{eq:BC2}
\end{align}
Importantly, one must ensure the hydrostatic pressure is continuous
across the boundary:
\begin{equation} \left. p(r) \right|_{R^-}^{R^+} = 0.
\end{equation}
The pressure for the Landau-Ginzburg functional is computed according
to:~\cite{CL_LG}
\begin{equation} \label{pressurefunctional} p(r) = -\cV + \sum_i \mu_i
  n_i + \sum_i \frac{\kappa_i}{2}\left(\frac{dn_i}{dr}\right)^2.
\end{equation}
In the stationary case, $\dot{R}=0$, the fluxes for each component
must be continuous as well:
\begin{equation} \label{eqflux}
  \left.  \left(\widetilde{D}_i \frac{\prtl
        \mu_i}{\prtl r}\right)\right|_{R^-}^{R^+} = 0
\end{equation}
Note we must separately specify the reaction rates at the boundary,
which we denote with $k_i^\ddagger$.

The above equations form a complete set of equations that allow one to
determine, self-consistently, the stationary value of the droplet
radius $R$. This setup is over-defined in the sense that not all
parameters are independent. Clearly, the reaction rates and
equilibrium concentrations are not independent:
\begin{equation} k_1^{(\alpha)} (n_{1,b}^{(\alpha)})^2 = 2
  k_2^{(\alpha)} n_{2,b}^{(\alpha)}
\end{equation}
for the 2nd order reaction and for the 1st order case:
\begin{equation}
  k_1 n_{1, b}^{(\alpha)} = k_2 n_{2, b}^{(\alpha)}.
\end{equation}
Here $\alpha$ labels the phase. A more subtle constraint on the
parameters comes about because of particle conservation. Indeed,
adding together the two equations in Eqs.~(\ref{particleCons_nonlin})
or (\ref{particleCons}) and using Eq.~(\ref{Fick}) yields
\begin{equation}
  \nabla^2 (\widetilde{D}_1 \mu_1 + \widetilde{D}_2 \mu_2) =
  0 \label{eq:total_diffusion}
\end{equation}
In equilibrium, this equation is solved by by $\widetilde{D}_1 \mu_1 +
\widetilde{D}_2 \mu_2 = 0$, since $\mu_i(r = \infty) = 0$ by
construction.  Thus one obtains
\begin{align} \label{eq:total_diffusion_balance}
  \widetilde{D}_1 \mu_1 = -\widetilde{D}_2 \mu_2.
\end{align}
Combining this with the boundary condition (\ref{eq:BC4}) yields
\begin{equation}
  \widetilde{D}_1^\text{(m)}/\widetilde{D}_2^\text{(m)} =
  \widetilde{D}_1^\text{(d)}/\widetilde{D}_2^\text{(d)}.
\end{equation}

\begin{figure}[t]
\centering
\includegraphics[width= \figurewidth]{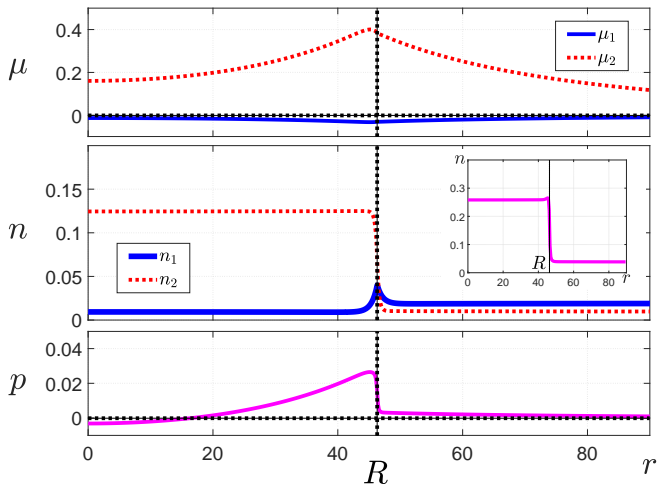}
\caption{\label{negp} The radial-coordinate dependences of the
  chemical potential $\mu_i$, concentrations $n_i$, and pressure $p$
  for a stationary cluster. The radial-coordinate, $r$, is measured
  from the center of the cluster. $\Delta g = 0.023$. The rest of
  the parameters are the same as Fig.~\ref{F:munpr_stat}.}
\end{figure}

For the reader's information, we illustrate in Fig.~\ref{negp} a
stationary droplet solution such that the pressure in the center of
the droplet is lower than in the bulk.

To approach non-stationary situations, we make additional
assumptions. First, we specify for concreteness that a droplet of the
minority phase is a vapor bubble with respect to the monomer, but a
liquid droplet with respect to the complex:
\begin{equation} \label{n1n2}
  \begin{array}{ll}  n_{1, b}^\text{(d)} &< n_{1, b}^\text{(m)} \\ \vspace{-2mm} \\
    n_{2, b}^\text{(d)} & > n_{2, b}^\text{(m)},
  \end{array}
\end{equation}
while assuming the monomer is the primary species in the majority phase:
\begin{equation} \label{n1n2out} n_{1, b}^\text{(m)} > n_{2,
    b}^\text{(m)}.
\end{equation}

Next we make the usual approximation~\cite{Bray} by which the
interface is assumed to move on timescales that are much longer the
diffusion times scales $R^2/D$. ($D$ is the regular diffusivity, see
below.) And so for each value of $R$, we solve the stationary
equations $\dot{n}_i^\ddagger = 0$ while relaxing the constraint
(\ref{eqflux}) that the fluxes of the components on the opposite sides
of the boundary be equal.  Using these assumptions, we (approximately)
infer the sign of the rate of change of the droplet radius away from
steady state:~\cite{Bray}
\begin{align} \dot{R} & \approx
  \frac{-1}{n^\text{(m)}_{1,b}-n^\text{(d)}_{1,b}} \left.
    \left(\widetilde{D}_1 \frac{\prtl \mu_1}{\prtl
        r}\right)\right|_{R^-}^{R^+} \nonumber \\ &=
  \frac{-1}{n^\text{(m)}_{2,b}-n^\text{(d)}_{2,b}} \left.
    \left(\widetilde{D}_2 \frac{\prtl \mu_2}{\prtl
        r}\right)\right|_{R^-}^{R^+}
  \label{eq:Rdot}
\end{align}
Note the second equation in Eq.~(\ref{eq:Rdot}) represents an
additional constraint.  Thus pegging $R$ and ${n}^\dagger_1$ away from
their stationary values allows one to find self-consistently to
determine the values of $\dot{R}$ and, for instance,
$\dot{n}^\dagger_1$. The corresponding flow chart is shown as
Fig.~\ref{F:flow} and demonstrates that the stationary solution in
fact represents a transition state, not a metastable configuration.

\subsection{First order reaction}
 
When the monomer-dimer conversion is a first order reaction, the
problem reduces to a set of two linear, fourth-order differential
equations, for each individual phase:
\begin{align}\begin{array}{ll} \label{ndotFinal} \dot{n}_1 &=
    \tilde{D}_1\nabla^2
    (-\kappa_1 \nabla^2 n_1 + m_1 n_1) - k_1 n_1 + k_2 n_2,
    \\ \vspace{-2mm} \\  \dot{n}_2 &= \tilde{D}_2\nabla^2 (-\kappa_2
    \nabla^2 n_2 + m_2 n_2) + k_1 n_1 - k_2 n_2.
\end{array}
\end{align}
subject to the the patching conditions discussed above and the
boundary conditions in the center of the droplet, $r=0$, and in the
bulk, $r = \infty$. In a standard fashion, we require that
\begin{equation} \label{bc} \begin{array}{rcl} n_i (r = \infty) &=&
    n_{i, b}
    \\     \\
    \bm \nabla n_i (r = 0) &=& 0.
    \end{array}  
\end{equation}
and
\begin{equation} \label{bcmu} \begin{array}{rcl} \mu_i (r = \infty)
    &=& 0  \\  \\
    \bm \nabla \mu_i (r = 0) &=& 0.
    \end{array}  
\end{equation}

\begin{figure}[t]
\centering
\includegraphics[width= .9 \figurewidth]{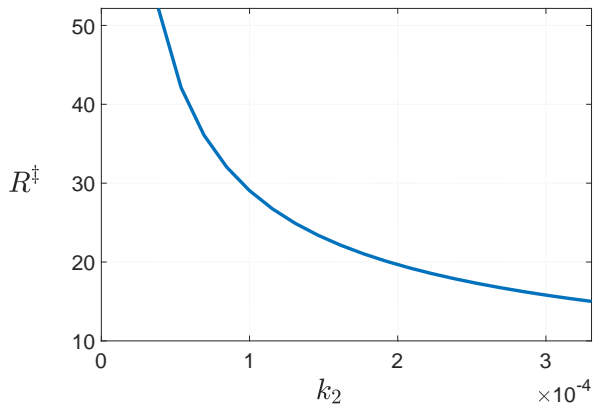}
\caption{\label{Rvsk} Dependence of the critical radius on the decay
  rate of the dimer $k_2 \equiv k^\text{(d)}_2 = k^\text{(m)}_2$. The
  rest of the parameter values are the same as in
  Fig.~\ref{F:munpr_stat}.}
\end{figure}

The linear equations~(\ref{ndotFinal}) are solved by a linear
superposition of Yukawa potential-like functions
$e^{qr}/r$.~\cite{CL_LG} The characteristic equation for the
wavevector $q$ can be written in a relatively transparent form:
\begin{align}
  0 &= q^6 - q^4\left(l_1^{-2} + l_2^{-2}\right) \nonumber\\
  &+ q^2\left[\left(l_1l_2\right)^{-2} + \left(l_1L_1\right)^{-2}
    + \left(l_2L_2\right)^{-2}\right] \nonumber\\
  &- \left[\left(l_1l_2L_1\right)^{-2} +
    \left(l_1l_2L_2\right)^{-2}\right], \label{eq:characteristic_length}
\end{align}
where $l^2_i \equiv \kappa_i/m_i$ and $L^2_i = D_i/k_i$. Here, 
\begin{equation} \label{DDm} D_i \equiv \widetilde D_i m_i
\end{equation}
is the ordinary diffusivity. Indeed, Eqs.~(\ref{eq:free_energy}) and
(\ref{Fick}) together with the usual $\bm j_i = - D_i \nabla n_i$ lead
to Eq.~(\ref{DDm}).  The lengths $l_i$ are, of course, the correlation
lengths of the Landau-Ginzburg theory;~\cite{Goldenfeld} they are
static, thermodynamic quantities. In contrast, the lengths $L_i$
originate exclusively from the presence of chemical conversion and are
{\em kinetic} quantities that constitute new length-scales in the
problem analogously to the length scale from
Eq.~(\ref{RDk}). Coefficients at the respective terms $e^{qr}/r$ are
constrained by the boundary and patching conditions, in the usual
way.~\cite{CL_LG} Cases when the characteristic roots are degenerate
can be dealt with straightforwardly. For instance, the doubly
degenerate root $q = 0$ corresponds to an additive constant and a
$1/r$ contribution to the overall solution. We observe that according
to Fig.~\ref{Rvsk}, the critical radius is largely determined by those
kinetic lengths.

\subsection{Second-order reaction}

As before, we solve exclusively for the stationary state within each
individual phase.  The stationary non-linear equations are solved
using the finite differences.~\cite{ascher1988numerical} We sub-divide
the space into three spherically-symmetric regions, all centered at
the origin: (1) the minority phase, $r < R$, (2) the vicinity of the
cluster in the majority phase, $R < r < R_p$, and (3) the outer
regions, $r > R_p$. The edge of the outer region, $R_p$, is chosen to
be sufficiently far away from the cluster boundary so that the
concentrations of the components are numerically close to their bulk
values. Thus in the outer region, the reaction-diffusion scheme can be
approximated by linearized equations in a controlled fashion:
\begin{align} \label{particleCons_linearize} \begin{array}{ll}
    \dot{n}_1 &= -\nabla \bm j_1 - 2 k_1 \delta n_1 n_{1,b} +
    2 k_2 \delta n_2 \\  \vspace{-2mm} \\
    \dot{n}_2 &= -\nabla \bm j_2 + k_1 \delta n_1 n_{1,b} - k_2 \delta
    n_2.
\end{array}
\end{align}
where $\delta n_i \equiv n_i-n_{i,b}$ is the deviation of
concentration of species $i$ from its bulk value.  The solution of the
linearized Eqs.~(\ref{particleCons_linearize}) is obtained exactly the
same way as the first order case from Eqs.~(\ref{particleCons}).

In regions 1 and 2, we solve the original non-linear equation using
finite differences while imposing patching conditions with the
linearized solution in region 3, at $R=R_p$. The patching is done by
enforcing that the density and the chemical potential of both species
be continuously differentiable at $r = R_p$.  The boundary conditions
at the cluster center, $r=0$, at the phase boundary, $r=R$, and in the
bulk, $r=\infty$ are the same as in the 1st order case.

\begin{figure}[t]
\centering
\includegraphics[width= 0.75\linewidth]{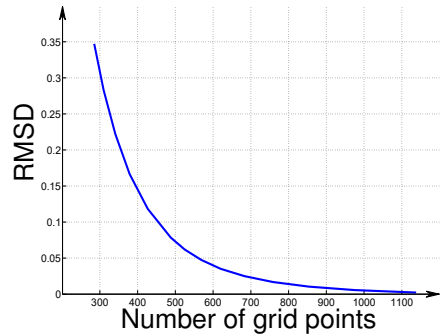}
\caption{\label{F:FigS3} 2nd-order reaction case: The root-mean-square
  difference of $\mu_1$ with a reference solution as a function of the
  number of grid points. The number of grid points of the reference
  solution is 1350.  $R_p = 85$. The rest of the parameter values are
  the same as in Fig.~\ref{F:munpr_stat_nonlin}.}
\end{figure}

\begin{figure}[t]
\centering
\includegraphics[width= 0.75\linewidth]{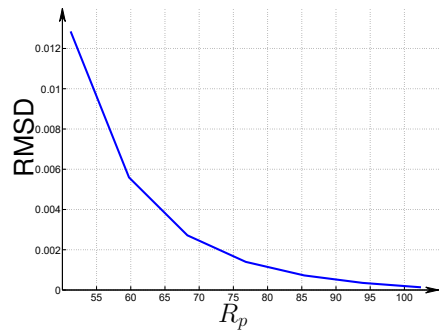}
\caption{\label{F:FigS4} 2nd-order reaction case: The root-mean-square
  difference of $\mu_1$ with a reference solution as a function of the
  patching radius $R_p$. In that reference solution, $R_p = 110$.  The
  number of grid points is fixed at 6 per unit length and The rest of
  the parameter values are the same as in
  Fig.~\ref{F:munpr_stat_nonlin}.}
\end{figure}

To test the convergence of our solutions, we compute them at several
values of the grid size and the patching radius $R_p$. We then
evaluate the root-mean-square (RMS) difference between these solutions
and the reference solution, which was obtained using some large number
of grid points and $R_p$ respectively. In Figs.~\ref{F:FigS3} and
\ref{F:FigS4}, we show the respective RMS differences for the chemical
potential of the monomer. These graphs indicate that our solutions do
in fact tend to a stationary value as the number of grid points and
$R_p$ are increased.

\subsection{Ripening}

Consider regular Ostwald ripening.  At a given value of over-saturation
$\Delta$ of the majority phase, the rate of growth of an individual
droplets is given by:~\cite{LifshitzSlyozov1961}
\begin{equation}\label{Rdotundersat}
\dot{R}=\frac{D}{R}\left(\Delta -\frac{\alpha}{R}\right),
\end{equation}
where $D$ is the diffusivity of the species in question and the
coefficient $\alpha$ is proportional to the mismatch penalty between
the majority and minority phases.~\cite{LLstat} The critical radius is
thus given by
\begin{equation}\label{smallcritR}
  R^\ddag = \alpha/\Delta.
\end{equation}
Eq.~(\ref{Rdotundersat}) can be profitably rewritten in terms of the
critical radius and the dimensionless radius $\widetilde R \equiv
R/R^\ddag$:
\begin{equation}\label{LSW_eq}
  \frac{d \widetilde R}{dt} = \frac{\alpha D}{{R^\ddag}^3} 
  \left(1-\frac{1}{\widetilde R}\right)\frac{1}{\widetilde R}
\end{equation}
Lifshitz and Slyozov~\cite{LifshitzSlyozov1961} have argued that at
sufficiently long times, the droplet size distribution tends toward a
scale-free form that is determined by the critical radius $R^\ddag$
alone. In other words, the distribution of the dimensionless radius
$\widetilde R$ is time independent at long times. Averaging
Eq.~(\ref{LSW_eq}) w.r.t. to this distribution immediately shows that
for this equation to be internally consistent, one must have at long
times:
\begin{equation}\label{LSWscaling}
R^\ddag = c (D\alpha t)^{1/3},
\end{equation}
where $c$ is a numerical constant of order one. (The constant turns
out to be 2/9 in the simplest treatment.~\cite{LifshitzSlyozov1961})
To avoid confusion, we note that the time are sufficiently long that
memory of the initial distribution of the droplet sizes is already
lost but not too long so that the number of clusters is still
sub-thermodynamic.  Eq.~(\ref{Rdotundersat}) implies that the
volumetric rate of droplet growth is linear in the quantity
$R-R^\ddagger$:
\begin{equation} \label{GT} R^2 \dot{R} \propto (R/R^\ddag-1).
\end{equation}

According to the discussion in the main text, our kinetically
stabilized clusters will exhibit ripening. Since they do not obey the
exact linear relation (\ref{GT}) we may inquire whether the ripening
exponent in the $R^\ddag$ vs. $t$ relation would differ significantly
from the value 1/3 from Eq.~(\ref{LSWscaling}) predicted by the
Lifshitz-Slyozov-Wagner theory and, in the first place, from the
experimental data due to Li et al.~\cite{LLVFPostwaldRipening} To
answer this question, we first fit the pertinent curve in
Fig.~\ref{F:Fig5} by a functional form:
\begin{equation} \label{eq:ripening_scale}
R^2 \dot{R} \propto R^x R^{\ddag^z} (R^y - R^{\ddag^y})
\end{equation}
Hereby, the Gibb-Thompson relation and diffusion-limited droplet
growth would correspond to $x = 0$, $y = 1$, and $z = -1$.) The same
line of logic that led to Eq.~(\ref{LSWscaling}) yields
\begin{equation} \label{eq:modified_LSW}
R^\ddag \propto t^{1/[3-(x+y+z)]}
\end{equation}

\begin{figure}[t]
\centering
\includegraphics[width= 0.75\linewidth]{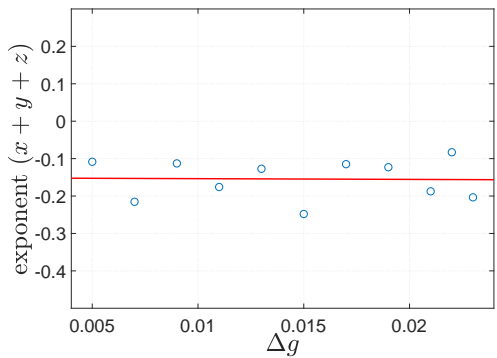}
\caption{\label{F:FigS1} The $\Delta g$ dependence of the combination
  $(x+y+z)$ of the parameters from Eq.~(\ref{eq:ripening_scale}) and
  (\ref{eq:modified_LSW}). $\Delta g$ is the bulk free energy excess
  of the minority phase per unit volume from Eq.~(\ref{Deltag}).}
\end{figure}

In Fig.~\ref{F:FigS1}, we show the $\Delta g$ dependence of the
combination $(x+y+z)$ of the parameters from
Eqs.~(\ref{eq:ripening_scale}) and (\ref{eq:modified_LSW}). We observe
that by Eq.~(\ref{eq:modified_LSW}), the predicted growth implies
$R^\ddag \propto t^{1/(3.1 \pm 0.1)} = t^{0.32 \pm 0.01}$, which is
quite close to both the experiment by Ye Li et
al.~\cite{LLVFPostwaldRipening} and the predictions due to the
Lifshitz-Slyozov-Wagner theory.~\cite{LifshitzSlyozov1961, WagnerOR,
  Bray, PhysRevA.20.595} We note that we have not shown that the
cluster-size distribution is, in fact, scale-invariant within the
present framework, which would be necessary to fully validate
Eq.~(\ref{eq:ripening_scale}). This is work in progress. Still,
experimental data due to Ye Li et al.~\cite{LLVFPostwaldRipening}
suggest that the distribution is, in fact, scale-invariant.

\end{document}